\documentclass[12pt]{article}
\usepackage{graphics}
\usepackage{graphicx}
\usepackage{amssymb}

\input{epsf}
\usepackage{cite}
\def\@fmsl@sh#1#2#3{\m@th\ooalign{$\hfil#1\mkern#2/\hfil$\crcr$#1#3$}}
 \def\eq#1\en{\begin{equation}#1\end{equation}}
\def\s[#1,#2]{[#1\stackrel{\star}{,}#2]}
\def\sx[#1,#2]{[#1\stackrel{\star_{x}}{,}#2]}

\newcommand{\nc}{\newcommand}
\nc{\beq}{\begin{equation}}
\nc{\eeq}{\end{equation}}
\nc{\beqa}{\begin{eqnarray}}
\nc{\eeqa}{\end{eqnarray}}

\def\bc{\begin{center}}
\def\ec{\end{center}}

\def\gsim{\mathrel{\rlap{\lower4pt\hbox{\hskip1pt$\sim$}}
    \raise1pt\hbox{$>$}}}       

\begin{document}
\makeatletter
\def\fmslash{\@ifnextchar[{\fmsl@sh}{\fmsl@sh[0mu]}}
\def\fmsl@sh[#1]#2{%
  \mathchoice
    {\@fmsl@sh\displaystyle{#1}{#2}}%
    {\@fmsl@sh\textstyle{#1}{#2}}%
    {\@fmsl@sh\scriptstyle{#1}{#2}}%
    {\@fmsl@sh\scriptscriptstyle{#1}{#2}}}
\def\@fmsl@sh#1#2#3{\m@th\ooalign{$\hfil#1\mkern#2/\hfil$\crcr$#1#3$}}
\makeatother


\title{\large{\bf Effective Theory for Dark Matter and a  New Force in the Dark Matter Sector}}

\author{Xavier~Calmet\thanks{Charg\'e de recherches du F.R.S.-FNRS} \thanks{xavier.calmet@uclouvain.be} and Swarup Kumar Majee \thanks{swarup.majee@uclouvain.be}\\
Catholic University of Louvain \\
Center for Particle Physics and Phenomenology\\
2, Chemin du Cyclotron\\
B-1348 Louvain-la-Neuve, Belgium
}

\date{May 2009}

\maketitle

\begin{abstract}
An effective theory for dark matter has recently been proposed. The key assumption is that the dark matter particle which is a Dirac fermion is protected from decaying by a global U(1) symmetry. We point out that quantum gravity effects will violate this symmetry and that the dark matter candidate thus decays very fast. In order to solve that problem, we propose to consider a local gauge symmetry which implies a new force in the dark matter sector. It is likely that this new local U(1) symmetry will need to be spontaneously broken leading for a range of the parameters of the model to a Sommerfeld enhancement of the annihilation cross-sections which is useful to explain the Pamela  and ATIC results using a weakly interacting massive particle with a mass in the TeV range.
\end{abstract}


\newpage

Pamela \cite{Adriani:2008zr} and ATIC \cite{:2008zzr} have recently reported 
excesses of positrons (to be precise electrons+positrons in the case of ATIC) which can be interpreted as a signal of Dark Matter although a local source of positron could also explain their signals. These observations have triggered a plethora of papers offering different interpretations within specific models. Most of these models are designed to address the issue linked with the stability of the Higgs boson's mass with respect to radiative corrections. This stability or hierarchy problem is however not based on mathematical or physical necessities but it rather linked to speculations on how gravity and the standard model are embedded into one fundamental theory. The Standard Model with a single Higgs boson, still to be discovered, does not suffer from any fundamental problem. The observations by Pamela and ATIC are very exciting because they suggest that new physics beyond the standard model is around the corner and as we shall see imply a dark matter particle with a mass of roughly 600 to 800 GeV. This is the strongest motivation to date to extend the Standard Model. We shall assume that the dark matter candidate is a Dirac fermion. However, this is clearly a theoretical prejudice as bosonic fields are viable dark matter candidates, see e.g.\cite{McDonald:1993ex,Hambye:2008bq}.

Since we do not want to limit ourself to a specific extension of the Standard Model, we shall follow the approach proposed by Harnik and Kribs  \cite{Harnik:2008uu} and consider an effective theory approach to dark matter.
It seems difficult to accommodate the Pamela and ATIC signals with standard extensions of the Standard Model and the idea of a hidden dark sector (such as e.g. \cite{Calmet:2003uj})  is growing in popularity.  
We want to reconsider an effective theory for a fermionic dark matter, a singlet under the standard model gauge group, which has been recently  proposed in ref. \cite{Harnik:2008uu}.  The key assumption is that the 
dark matter particle $D$ is protected from decaying by a global U(1)  symmetry. Allowing four-fermion leptonic operators only the  authors of    \cite{Harnik:2008uu} have  succeeded to explain Pamela data for a Dirac fermion dark matter with 
mass of the order of electroweak scale with no or a minimal boost.  
 Discussing briefly the effective theory of Dirac dark matter
 with a global symmetry, in this article we show different ways in which 
the Dirac fermion decays to other light particles and it's dramatic impacts 
on the cosmological evolution of the density of dark matter. 

We shall start by reviewing the proposal of Harnik and Kribs  \cite{Harnik:2008uu}. A Dirac fermion $D$ is assumed to be stable under
some global symmetry $U(1)$. This global symmetry prevents  the fermion $D$ from decaying and it can only annihilate to the standard model particle through some higher-dimensional operators suppressed by a high energy scale $\Lambda$. The dark matter particle is assumed to be a weakly interacting massive particle (wimp). The largest
annihilation cross section will appear in the annihilation mode to a  pair of Higgs scalars via dimension-five operators
\begin{eqnarray}
\frac{ \bar{D} D H^\dagger H}{\Lambda}.
\end{eqnarray}
However, the effect of this operator, as described by Harnik and Kribs is model dependent and for a Dirac bino, for example, would be irrelevant.  On the other hand a relatively weaker, namely, dimension-six operators, 
for example, 
\begin{eqnarray}
\frac{\bar{D}\gamma^{\mu} D\bar {f}_{L/R} \gamma_{\mu} {f_{L/R}}}{\Lambda^2},
\end{eqnarray}
will allow the dark matter particles to directly annihilate into left/right chiral 
fermions $f_{L/R}$. Negligible small cross-sections of the direct detection of 
spin-independent scattering interaction implies the absence of any interaction
to the quarks \cite{Ahmed:2008eu,Angle:2007uj}. This leaves the dark 
matters to annihilate to leptons only. If one considers in a supersymmetric 
theory  the Dirac bino as the dark matter candidate it would then 
strengthen the right-handed leptonic decay channel and it  would thus explain the 
observed spectrum by Pamela. 

The proposal of Harnik and Kribs is very interesting, however it is expected that global symmetries cannot be exact symmetries of Nature. Indeed, it is expected that quantum gravity violates global symmetries, see for example \cite{Holman:1992us} via e.g. virtual quantum black holes.  Even if the breaking of the global U(1) is small, this can have a dramatic impact for the effective theory of dark matter considered above. Let us classify the operators which can be generated by quantum gravity effects. The lowest dimension operator compatible with all the gauge symmetries of the Standard Model is of the type
\begin{eqnarray}
\epsilon \bar L H D_R,
\end{eqnarray}
where $\epsilon$ is a numerical coefficient, $L$ 
is a left-handed SU(2) lepton doublet,   $H$ is the Higgs doublet of the 
standard model and $D_R$ is the right-handed dark matter particle. These operators will be generated since they are compatible with all the 
symmetries of the standard model. However,  it is plausible that dimension four operators generated by quantum gravity are suppressed by a factor $\exp(-\lambda/M_P)$ where $M_P$ is the Planck mass and $\lambda$ could be some low energy scale since in the limit $M_P \to \infty$ the operator must disappear. But, the suppression has to be huge.
Unless the numerical coefficient $\epsilon$ 
is minuscule these operators will lead to a decay of the dark matter particles long before our time. An estimate of its lifetime is given by $\tau \sim 8 \pi (\epsilon M_D)^{-1}$. The Higgs boson can be on or off-shell. In a wimp scenario, one expects $M_D \sim 100$ GeV. This operator leads to a lifetime for the dark matter candidate much short than that of the universe unless $\epsilon < 8 \times 10^{-43}$. 

The next operator of dimension 5 which we wish to consider is 
\begin{eqnarray}
\frac{\epsilon_1}{M_P} \bar D  H^\dagger \fmslash D L
\end{eqnarray}
where $M_P$ is the Planck mass and $\fmslash D$ is the covariant derivative of the standard model which transforms in the adjoint of SU(2). This class of operators leads to a dimension four operator after symmetry breaking:
\begin{eqnarray}
\sim \frac{\epsilon_1 v}{M_P} \bar D   \fmslash Z  \nu_L, 
\end{eqnarray}
where $v$ is the Higgs vacuum expectation value. Clearly  the suppression of this operator is not sufficient to protect the dark matter candidate to have completely annihilated already at our cosmological epoch.  As in the previous case, the Z boson can be on-shell or off-shell. An estimate of its decay width is given by $\Gamma \sim M_D/8\pi (\epsilon_1 v/M_P)^2$ where $v$=246 GeV. For $\epsilon_1 \sim 1$, one finds a lifetime of the order of  $\tau \sim M_D^{-1} \times 10^{-32}\sim 10^6$ s which is obviously much shorter than the age of the universe $\sim 3 \times 10^{17}$ s.

If there are right-handed neutrinos in nature which are gauge singlets under the standard model, one can consider other  dangerous operators, for example 
\begin{eqnarray}
\frac{\epsilon_2}{M_P} \bar D_L \nu_R H^\dagger H,
\end{eqnarray}
 where  $M_P$ is the Planck mass and $\nu_R$ is a right-handed neutrino.  This class of operators also leads to a rapid decay of the dark matter candidate. Again, the Higgs boson can be on-shell or off-shell. 

We propose to gauge the global U(1) symmetry to solve this problem. Since quantum gravity preserves gauge symmetries, gauging this symmetry will prevent the operators discussed above from appearing at the quantum level. Interestingly effective theory arguments lead to the conclusion that there is a fifth force in Nature acting in the dark matter sector. This idea has been discussed intensively in the literature, see e.g \cite{Ackerman:2008gi,Farrar:2006tb}.   A new dimension four operator mixing the hidden sector and the standard model is allowed:
\begin{eqnarray}
\epsilon_3 F_{\mu\nu} \hat F^{\mu\nu} 
\end{eqnarray}
where $\hat F^{\mu\nu}$ is the field-strength of the new local U(1) symmetry. This operator leads to a mixing between the standard model hyperphoton and the new photon, usually called paraphoton (see e.g. \cite{Dobrescu:2004wz}) for a recent review).  

There are now different possibilities. The hidden sector gauge symmetry could be unbroken and Ackerman et al. \cite{Ackerman:2008gi} have shown that if the dark matter field is a singlet under the standard model gauge group and it forms the thermal relic, then it would violate bounds coming from limits on hard and soft scattering for many values of the parameters of the model. However if it is gauged under $SU(2)_L$ the bounds are relaxed. This goes against the spirit of the effective theory we are considering which is based on the idea that the dark matter is part of a hidden sector. The natural alternative is to break the extra U(1) spontaneously. We thus have to introduce a new Higgs boson $\phi$ in the hidden sector. This corresponds to the case discussed recently by Arkani-Hamed et al. \cite{ArkaniHamed:2008qn}. Besides the mixing of the photon and the paraphoton, there is a new renormalizable interaction linking the two Higgs sectors through an operator $\alpha \phi^* \phi H^\dagger H$. This coupling could lead to missing energy signals at the LHC \cite{Calmet:2003uj,Calmet:2006hs}.  Another motivation to break this symmetry is to generate a Sommerfeld enhancement which as pointed out in \cite{ArkaniHamed:2008qn} could lead to an explanation of both the Pamela and ATIC excesses if the mass of the dark matter fermion is in the 500-800 GeV range (which is fixed by the ATIC excess) and the mass of the massive paraphoton is in the few GeV range (which is fixed by the Pamela excess). We shall not review the phenomenology of that model which can be found in  \cite{ArkaniHamed:2008qn}.

Note that in the limit where the symmetry breaking scale of the hidden U(1) is bigger than the Planck mass, the global symmetry becomes a Z$_2$ symmetry which is conserved by quantum gravity since it originates from a local gauge symmetry. This resembles very much the global symmetry discussed at the beginning of this work and it would be impossible experimentally to differentiate the global U(1) from a Z$_2$ symmetry. Nevertheless at the fundamental level we would know that there is a new force in the hidden dark matter sector which is even weaker than gravity.

There is to date no real theory of dark matter and the motivations to study physics beyond the standard model are strongly dependent of theoretical prejudices. It is thus important to remain as model independent as possible. In this note we consider a hidden dark matter sector and classify the possible couplings to the standard model. We show that quantum gravity would lead to a breaking  of any global symmetry introduced in the hidden sector to prevent the dark matter particle from decaying. An obvious way out of this problem is to gauge this symmetry. This has deep consequences as it implies the existence of a new force in Nature. It is interesting that this follows from arguments based on effective field theory only. If this new dark force is broken, one could easily explain the excess of ATIC and Pamela using the wimp mechanism and a Sommerfeld enhancement which can appear naturally for a wide range of parameters. If, however, the hidden gauge symmetry is unbroken, there is range of  parameters for which the dark matter could lead to the correct abundance. The price to pay would be a fine-tuning of the parameters as pointed out in \cite{Ackerman:2008gi}. More observational and experimental data are required to determined the parameters of the effective theory described above.

\subsection*{Acknowledgments}
We would like to thank Stephen Hsu and Graham Kribs for useful communications. This work is supported in part by the Belgian Federal Office for Scientific, Technical and Cultural Affairs through the Interuniversity Attraction Pole P6/11.

\noindent


\bigskip


\end{document}